\documentclass[twocolumn,amsmath,amssymb,prb]{revtex4}

\usepackage{graphicx}
\usepackage{dcolumn}
\usepackage{bm}

\begin{document}

\title{Thermal spin transport and spin-orbit interaction in ferromagnetic/non-magnetic metals}

\author{Abraham Slachter}
 \email{A.Slachter@gmail.com}
\author{Frank Lennart Bakker}
\author{Bart Jan van Wees}%

\affiliation{Physics of Nanodevices, Zernike Institute for Advanced Materials, University
of Groningen, The Netherlands}

\date{\today}

\begin{abstract}
In this article we extend the currently established diffusion theory of spin-dependent electrical conduction by including spin-dependent thermoelectricity and thermal transport. Using this theory, we propose new experiments aimed at demonstrating novel effects such as the spin-Peltier effect, the reciprocal of the recently demonstrated thermally driven spin injection, as well as the magnetic heat valve. We use finite-element methods to model specific devices in literature to demonstrate our theory. Spin-orbit effects such as anomalous-Hall, -Nernst, anisotropic magnetoresistance and spin-Hall are also included in this model. \end{abstract}

\maketitle

\section{Introduction}
Spintronics uses the spin degree of freedom to demonstrate new functionality in ferromagnetic/non-magnetic hybrid devices\cite{Zutic}. In time, many new functional devices have been proposed\cite{Slonczewski,Brataasspinpump,Hirsch} and measured\cite{Kiselev,Costachespinpump,Saitoh,Valenzuela} utilizing the special properties of spin transport. Recently, the coupling between thermoelectricity and spin transport has been added to this field. New applications resulting from this coupling are summarized under the branch called spin-caloritronics\cite{BauerSpinCaloritronics,Heikkila,Saitoh,Jaworski,Uchida2,Jansen,Tulaparkur2,Giazotto}.

A diffusive transport theory for spin-dependent electrical conduction is currently well established\cite{Mott,Valet}. This theory has been extended to non-collinear systems\cite{BrataasNCME,Tserkovnyakrevmodphys} which becomes relevant when spin-dependent tunneling through interfaces is considered or to quantify dynamic processes such as spin-transfer torque\cite{Chappert} or spin-pumping\cite{Tserkovnyakspinpump,Brataasspinpump}.

In this article, we extend the collinear theory of diffusive transport for spin-dependent conduction to include spin-dependent thermoelectricity\cite{Johnson3,Wegrowe,Gravier2}, spin-orbit effects\cite{Valenzuela,Kimura,Slachter2} and also spin-dependent thermal transport. We use finite-element methods to demonstrate our theory in order to  extract useful parameters from complex three-dimensional device geometries\cite{Bakker,Slachter}. Various recent experiments are taken from literature to extract the parameters which govern the effects.

The setup of this article is as follows. In section II, we begin with a description of finite-element modeling where we specify the structure of the model and the solvers used. In Section III we describe how to make finite-element-models which describe electrical spin-transport. We illustrate this model by calculating a recent example from literature\cite{Yang}. We also show how the direct and inverse spin-Hall effect can be included and use it to model an experiment by Kimura et al.\cite{Kimura}. A thermoelectric model which excludes spin is described in section IV where it is shown how spin-orbit effects can be included\cite{Bakker,Slachter2}.

In section V, we introduce the thermoelectric-spin model. This model can describe the individual effects related to thermoelectricity or spin-dependent electrical transport. However, the introduction of spin-dependent thermoelectric coefficients also allows to demonstrate two new physical effects: the recently demonstrated thermal spin injection\cite{Slachter} and its Onsager reciprocal effect: the spin-Peltier effect. Thermal spin injection describes the injection of spins in a non-magnetic material when a heat current is sent through a ferromagnetic/non-magnetic interface. The spin-Peltier effect describes spin-dependent heat transport across this interface due to the injection of spins in the ferromagnetic material.

In section VI, a phenomenological theory for spin-dependent heat transport is proposed, where the concept of a spin temperature is introduced and the thermal analogy of the spin valve: the magnetic heat valve\cite{Heikkila}. We apply the model on a previously measured sample\cite{Slachter2} to determine an upper limit for the relaxation of spin-dependent heat at room temperature. Thermoelectricity not only connects spin-dependent charge transport to heat transport but also connects spin-dependent heat transport to charge transport. This provides new ways to generate spin temperatures and to detect these.

We conclude this article with a discussion on how the spin-dependent effects can be described by a theory which includes tunneling through interfaces or non-collinear magnetizations\cite{Hatami}.

\section{Finite-element modeling}

The finite-element modeling in this article is performed using the software package Comsol Multiphysics (version 3.5). It solves partial differential equations (PDE's) for 1, 2 or 3 dimensional geometries defined in a CAD drawing program. In a diffusive transport theory, the PDE's are determined by the conservation of the generalized currents for the physics considered. These can be formally derived from Boltzmann transport theory\cite{Valet,Zhang}. The fluxes, put into a vector by $\vec J = \left(J_{u_{1}},J_{u_{2}},...\right)$, are governed by a vector of continuous variables $\vec u = \left(u_{1},u_{2},...\right)$ through the conductance matrix $\bar c$:

\begin{equation}\label{eq2:Comsol1}
\vec J = - \bar c ~\nabla \vec u
\end{equation}

Depending on the dimensionality \textit{n} (1D, 2D or 3D) of the finite-element model, the elements of the fluxes $J_{u_{1}},J_{u_{2}},...$ are vectors themselves of size \textit{n}. They determine the currents in the respective directions defined by the coordinate system of the model. The elements of the conductance matrix $\bar c_(i,j)$ are then $n\times n$ matrices. For an isotropic model, these are scalar matrices while for anisotropic transport the elements can be different. The PDE's in the bulk are determined by the conservation of fluxes:

\begin{equation}\label{eq2:Comsol2}
\vec \nabla \cdot \vec J = \vec f\left(\vec u\right)
\end{equation}

\noindent Where a source term $\vec f\left(\vec u\right)$ exists which may depend on the variables themselves. As an example, for simple electrical transport $\vec u=V$, $\vec J = J_{c}$, $\bar c=\sigma$ and $f$=0. Here $V$ is the voltage, $J_{c}$ the charge current and $\sigma$ the electrical conductivity. Eq. \ref{eq2:Comsol1} then states Ohm's law while Eq. \ref{eq2:Comsol2} is the Poisson equation representing the conservation of charge. The system under consideration is solved by stating the boundary conditions. These can be set for each variable (a Dirichlet condition) or flux (a Neumann condition) individually. In our example of electrical conduction, a charge current can be sent through the material by setting the charge current to a specific value at one interface and the voltage to a specific value at another. The outer interfaces are insulating $J_{c}$=0 and the currents are continuous across internal interface $J_{c}|_{1}$ = $J_{c}|_{2}$.

A (tetrahedral) mesh of typical 300k elements is created by the finite-element program where specific detailed meshing is often used in the areas of interest by specifying a minimal element size. The PDE's are solved using a built-in (non-linear) solver which uses the iterative generalized minimal residual solver (FGMRES) with a geometric multigrid preconditioner, which in its turn uses a a direct sparse object-oriented linear equations solver (SPOOLES).

The models we use are generally non-linear and the device on which the model is based is measured electrically. Therefore, the resulting measurable voltage is also non-linear:

\begin{equation}\label{eq2:R1R2factors}
V=R_{1}I+R_{2}I^{2}+R_{3}I^{3}+...
\end{equation}

The contributions $R_{n} (V/A^{n})$ to the nonlinear voltage can be separately determined from experiments. By using multiple lock-in systems which are set to measure the different responses $\omega$, 2$\omega$, 3$\omega$ resulting from a sinusoidal charge current $I$ of frequency $\omega$ sent through the device, it is possible to determine these contributions\cite{Slachter}. We may extract them from the constructed model by studying how the simulated voltages depend on the applied charge current. The nonlinear contributions $R_{n} (V/A^{n})$ are determined by calculating the model at currents of $\pm I$ and $\pm 2I$ and deriving these contributions. Here $I$ is typically of a size used in experiments where all interesting contributions $R_{n} (V/A^{n})$ are considerable.

\section{Spin transport}
Spin-dependent electron transport in systems consisting of collinear magnetizations and clean ferromagnetic/non-magnetic interfaces is commonly described in terms of a 2-channel model. First suggested by Mott\cite{Mott}, and later derived from the Boltzmann transport theory\cite{Valet}, it describes electrical conductance separately for spin-up ($\uparrow$) electrons, the component parallel to a magnetization, and spin-down electrons ($\downarrow$), the antiparallel component. Each channel has its own conductivity $\sigma_{\uparrow,\downarrow}$, voltage $V_{\uparrow,\downarrow}$ and charge current $J_{\uparrow,\downarrow}$. Usually a simplified resistor model is employed to describe spin-dependent transport. While it is sufficient for many approximations, it can become inaccurate for complex three-dimensional structures\cite{Kimura2}.

Spin-dependent transport can be modeled by a set of PDE's by using the spin-dependent voltages as variables $\vec u$ = $\left(V_{\uparrow},V_{\downarrow}\right)$. The spin dependent currents $\vec J$ = $\left(J_{\uparrow},J_{\downarrow}\right)$ are then defined by the spin-dependent conductance matrix:

\begin{equation}\label{eq2:spinconductance}
\bar c = \left(\begin{array}{cc} \sigma_{\uparrow} & 0 \\ 0 & \sigma_{\downarrow}\end{array}\right)
\end{equation}

The conservation of charge current is given by $\vec \nabla\cdot\left(\vec J_{\uparrow}+\vec J_{\downarrow}\right)=0$. The Valet-Fert equation $\nabla^{2} (V_{\uparrow}-V_{\downarrow}) = \frac{V_{\uparrow}-V_{\downarrow}}{\lambda^{2}}$ is derived from the conservation of spin currents. Defining a spin polarization for electrical conductance P$_{I} = (\sigma_{\uparrow}-\sigma_{\downarrow}) / (\sigma_{\uparrow}+\sigma_{\downarrow})$ we find a source term:

\begin{equation}\label{eq2:sourceterm2channelmodel}
\vec f=\frac{(1-P_{I}^{2})\sigma}{4\lambda^{2}}(V_{\uparrow}-V_{\downarrow})\cdot\left(\begin{array}{c} -1 \\ 1\end{array}\right).
\end{equation}

\begin{figure}[t]
\includegraphics[width=8.8cm,keepaspectratio=true]{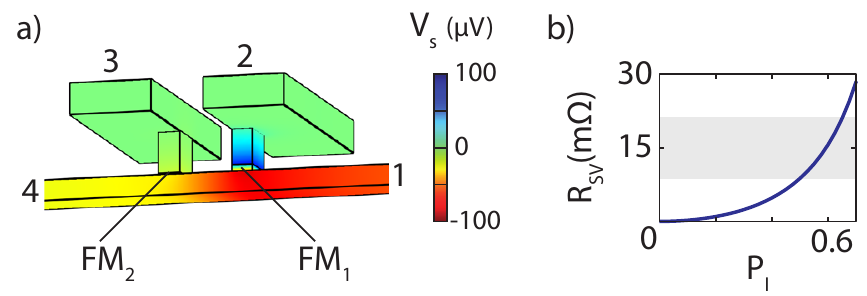}
\caption{\label{fig2:1} (Color online) Results of the modeling of the non-local spin valve structure used by Yang\cite{Yang} et al. a) A spin current is injected into the Cu bar which connects two ferromagnets FM$_{1}$, FM$_{2}$ by sending a charge current I$_{1-2}$ = 1 mA. The injected spins diffuse towards FM$_{2}$ where they are absorbed. The color shows a calculation of spin voltage $V_{s}=V_{\uparrow}-V_{\downarrow}$ at I$_{1-2}$ = 1 mA for the structure with a 4 nm thin ferromagnet FM$_{2}$.  b) Calculated spin valve signal V$_{3-4}$/I$_{1-2}$ (m$\Omega$) versus the spin polarization for electrical conductance for a 20 nm thick FM$_{2}$. The grey area shows the measured spread.}
\end{figure}

The inputs for this model are the specific geometry, the conductivity $\sigma=\sigma_{\uparrow}+\sigma_{\downarrow}$, spin relaxation length $\lambda$ and spin polarization $P_{I}$. These parameters can be determined from various experiments. For example, the relaxation length of non-magnetic materials can be determined by varying the distance between two ferromagnet materials in a spin valve or measuring spin precession\cite{Jedema,Jedema2}. The relaxation length of ferromagnets can be determined by angle-resolved photoemission\cite{Petrovykh}, while the conductance polarization of ferromagnets can be investigated by measuring the Doppler shift of spin-waves\cite{Vlaminck}.

While the model typically works well for pillar structures with clean interfaces\cite{Yang}, in lateral systems the model generally overestimates the observable spin-valve signals\cite{Kimura3}. This is because it is often necessary to perform ion-milling prior to deposition to obtain ohmic interfaces. This is taken into account in the modeling by using a reduced conductance polarization of the ferromagnet P$_{I}$. For example, for a permalloy (Ni$_{80}$Fe$_{20}$) ferromagnet commonly used in a lateral spin valve, the polarization is reduced from 70\%, determined from measured Doppler shifts, to 30-50\%\cite{Jedema,Kimura3}.

An example of the application of this model can be found in Bakker\cite{Bakker}. Another example of where this model can be applied is shown by Yang\cite{Yang} et al. Here, the non-local spin valve geometry\cite{Johnson,Jedema} is used to inject a spin current $ J_{\uparrow}-J_{\downarrow}$ free of charge current (a so-called pure spin current) into a thin ferromagnet. The magnetization of this magnet is switched by the resulting spin-transfer torque\cite{Chappert,Slonczewski}. A threshold exists for this process given in terms of the charge current which should be sent through the device.

A model of this device geometry with resulting spin voltage is shown in Fig. \ref{fig2:1}. Using the measured conductances, relaxation length of copper $\lambda_{Cu}$ = 1$\mu$m and spin valve signals, we determine an effective spin polarization of P$_{I}\approx$0.6 from the measured spin valve signals 9-21 m$\Omega$ from a batch of samples with thick ferromagnets. By performing an integration of the spin current flowing through the FM$_{2}$/NM interface, we find that the amount of pure spin current injected into the second ferromagnet I$_{s}$ is 13\% of the total charge current $I$ sent through the device. Using an effective formula for spin-transfer torque switching\cite{Sun} we find that the ferromagnet should switch at I$_{s}$=930$\mu$A using common parameters\cite{Costachespinpump} where in the experiment the required charge current I=5mA results in a a spin current of I$_{s}$=675$\mu$A. Considering the empirical spread found for spin valve signals, this is very reasonable.

Because the electrical current density spread throughout the device is modeled, the Biot-Savart law also allows to calculate the magnetic fields present in the device. The magnetic fields at position $\vec r$ is determined by performing a volume integral over the entire device:

\begin{equation}\label{eq2:BiotSavart}
\vec B (\vec r_{0})= \frac{\mu_{0}}{4\pi}\int\frac{\vec J_{c}\times(\vec r-\vec r_{0})}{|\vec r-\vec r_{0}|^{3}} d\vec r
\end{equation}

\noindent Here $\mu_{0}$ is the magnetic vacuum permeability. We find that the magnetic field at the center of the switchable ferromagnet is $|\vec B|$ = 1.2 mT at the maximum applied charge current of 5 mA while the switching field of the magnet is 8 mT. This directly rules out that the magnetization switches by the induced magnetic fields, but it can be responsible for observed asymmetries in the spin-transfer torque switching process.

\subsection{Spin-Hall Effect}

Spin-orbit effects in ferromagnets are often sizeable due to their complex band alignment. As a result, measurements on spin-orbit effects in ferromagnets were already reported more then a century ago. First D'Yakanov and Perel\cite{DYakanov} and later Hirsch\cite{Hirsch} suggested that the same process which governed these effects in ferromagnets, whether it be due to band alignment (intrinsic) or spin-dependent scattering (extrinsic), can also be responsible for a new effect in paramagnetic materials: the spin-Hall effect\cite{Kato,Valenzuela}.

The direct spin-Hall effect describes that when a charge current $J_{c}$ is sent through a material with strong spin-orbit interaction, a spin current $J_{s}$ flows away from the center of the conductor with its spin direction $\vec m$, a unity vector, perpendicular to the charge and spin current. In the indirect version, a spin current flowing through the material creates a voltage perpendicular to the spin and spin current direction. Based on a Boltzmann transport theory derived by Zhang\cite{Zhang}, the effects are governed by the following two equations:

\begin{align}\label{eq2:spinhall}
\vec\nabla V^{SH}&=-\frac{\theta_{SH}}{\sigma} \vec m\times \vec J_{s}\\
\vec\nabla V_{s}^{ISH}&=\frac{\theta_{SH}}{\sigma} \vec m\times \vec J_{c}
\end{align}

\begin{figure}[t]
\includegraphics[width=8.8cm,keepaspectratio=true]{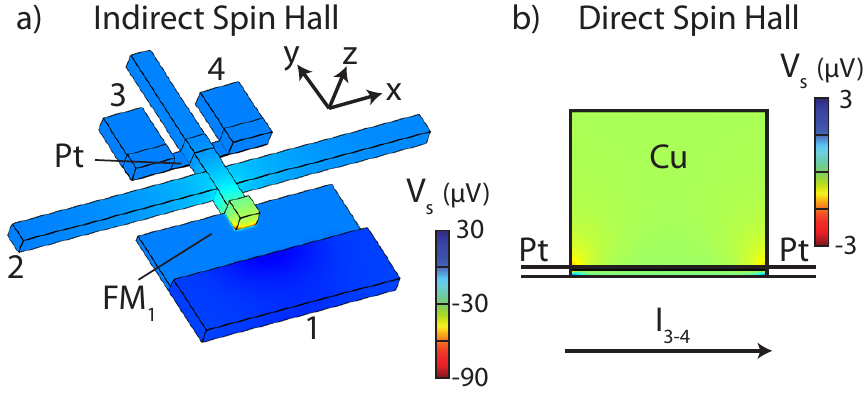}
\caption{\label{fig2:2} (Color online) Calculated spin voltages V$_{s}$ ($\mu$V) for the device of Kimura\cite{Kimura} et al. a) In the direct spin-Hall configurations a charge current I$_{1-2}$ = 1 mA is sent from ferromagnet to copper arm and the resulting spin-Hall voltage V$_{3-4}$ is measured on the Pt strip. b) XZ cross section in the middle of the platinum strip in the indirect spin-Hall configuration. The charge current I$_{3-4}$ = 50 $\mu$A is short-circuited by the copper strip which is why most spin accumulation enters through the corners. When the spin voltage enters the copper strip it is only a small fraction of the $\pm 8 \mu$V spin voltage present in the bulk platinum. The voltage V$_{1-2}$ is now measured by converting the spin voltage V$_{s}$ to a charge voltage at the FM/NM interface.}
\end{figure}

\noindent Here $\vec\nabla V^{SH}$ and $\vec\nabla V_{s}^{ISH}$ are the bulk charge and spin voltages resulting from the direct and indirect spin-Hall effect and $\theta_{SH}$ is the spin-Hall angle, typically a small fraction of one. Both effects can be included into the 2-channel model. To do this, we rewrite both equations into spin-up and spin-down currents and obtain the new spin-dependent conductance matrix:

\begin{equation}\label{eq2:spinhallmatrix}
\bar c = \left(\begin{array}{cc} \sigma_{\uparrow} & \sigma_{\uparrow}^{SH} \\ \sigma_{\downarrow}^{SH} & \sigma_{\downarrow}\end{array}\right)
\end{equation}

\noindent Where non-diagonal elements $\sigma_{\uparrow,\downarrow}^{SH}$ are included. These become skew-symmetric matrices determined by the spin-direction $\vec m$ considered in the device:

\begin{equation}\label{eq2:spinhallnondiagonallelements}
\sigma_{\uparrow,\downarrow}^{SH}(i,j)=\mp \theta_{SH} \sigma_{\uparrow,\downarrow}\sum_{k}\varepsilon_{ijk}m_{k}
\end{equation}

\noindent Here (i,j,k) are the indices of the predefined xyz axes and $\varepsilon_{ijk}$ is the Cevi-Levita symbol.

To demonstrate the theory, we model a device measured by Kimura\cite{Kimura} et al. where both the direct and indirect spin-Hall effect were measured in a single nanoscale device at room temperature for the first time. The results from this model are shown in Fig. \ref{fig2:2}. The device consists of a single permalloy ferromagnet which is connected to a 4 nm thin platinum strip by a copper cross. A pure spin current can be injected into the platinum strip by sending a charge current I$_{1-2}$ from the ferromagnet to one of the arms of the copper cross. When the magnetization of the ferromagnet is aligned in the $\pm$y-direction, the spin current flowing into the platinum in the -z direction creates a voltage due to the indirect spin-Hall effect in the x-direction. This voltage can be measured between the two contacts present on the platinum strip. In the same device, sending a charge current through the platinum strip creates a spin current flowing in the z-direction where the spins are aligned in the $\pm$y-direction, which is now due to the direct spin-Hall effect. When this spin current arrives at the permalloy strip, it is converted into a voltage which can be electrically measured.

Both the direct and indirect spin-Hall signal in this device is around 60 $\mu\Omega$ at room temperature at a distance of 400 nm from platinum strip to ferromagnet. Using the common parameters P$_{Py}$=0.3, $\lambda_{Py}$ = 5 nm, $\lambda_{Pt}$ = 2 nm and $\lambda_{Cu}$ = 350 nm\cite{Jedema,Kimura2,Kimura3} and the measured conductivities, we find that a spin-Hall angle of $\theta_{SH}$ = 5$\cdot10^{-2}$ accurately produces both signals. This angle is around 8 times smaller then previously deduced. The difference lies somewhat in the used parameters, which are different, but mostly in the modeling which takes into account effects which the resistor model does not. For example, when a charge current is sent through the contacts connected to the platinum strip, the charge current mostly goes through the copper on top of the platinum strip instead of the platinum strip itself, which is due to the high conductivity of the copper. Copper is expected to have virtually no spin-Hall effect which reduced the observed voltage. We calculate that only 1\% of the charge current goes fully through the platinum strip. A similar effect also occurs for the indirect spin-Hall effect. When a spin current is injected into the platinum strip, the produced voltage is also short-circuited by the copper strip. This example clearly demonstrates the need for a three-dimensional model to accurately predict the relevant parameters.

\section{Thermoelectricity}

Ferromagnetism of typical (metallic) ferromagnets used in spintronic experiments originates from the electronic band structure. This band structure varies more strongly than those of the non-magnetic parts due to ferromagnetically induced shifts in the s- and d-bands. The Mott-formula\cite{Cutler} describes that thermoelectricity is strongest when the electrical conductivity varies strongly around the Fermi energy. Since this variation is dependent on the band structure, the ferromagnets used in spintronic experiments are good thermoelectric materials\cite{Barnard}. A model of thermoelectricity is therefore highly required to explain the voltages observed in nanoscale devices\cite{Bakker,Slachter2} and can also be used to demonstrate new physics\cite{Slachter}.

A bulk thermoelectric theory based on Boltzmann transport has been in use for half a century\cite{Rowe,Ashcroft}. In this theory, in addition to charge transport $J_{c}$, the heat transport $Q$ [W/m$^2$] is also taken into account. In a thermoelectric model, we use the variables $\vec u = (V,T)$ and the currents $\vec J = \left(J_{c},Q\right)$. The transport is then determined by the conductance matrix:

\begin{equation}\label{eq2:Thermoelectricmatrix}
\bar c = \left(\begin{array}{cc} \sigma & \sigma S \\ \sigma \Pi & k\end{array}\right)
\end{equation}

\noindent where $k$ is the thermal conductivity, $S$ the Seebeck coefficient and $\Pi$ the Peltier coefficient. The Peltier coefficient is related to the Seebeck coefficient by the thermodynamic Thomson-Kelvin relation $\Pi = S T$, where T is the reference temperature. The Seebeck coefficient governs the Seebeck effect: the generation of a voltage as a result of a temperature gradient. The Peltier coefficient describes the transport of energy when a charge current is sent through a material.

In the model, the charge current remains conserved. However, the heat current is not. When a charge current is sent through a material, the inelastic scattering of electrons causes a rise in temperature. This is a process known as Joule heating. Conservation of the currents gives the source term: $\vec f = (0,\vec J_{c}^{2}/\sigma)$.

Unlike electrical transport, which is only carried by (missing) electrons, thermal conductivity typically has two contributions: that from the energy transported by electrons and from lattice vibrations: phonons. In metals, the transport by electrons often dominates. In this case, the Wiedemann-Franz law $k=\sigma L T$, which relates thermal and electrical conductivity through the Lorenz number $L$, can be used to estimate thermal conductivities from measured electrical conductivities.

This model has recently been demonstrated in nanoscale spintronic devices\cite{Bakker,Slachter2}. These experiments illustrate that the ferromagnetic/non-magnetic junction used acts as a thermocouple to measure the local temperature of the device. In addition, when a charge current is sent through such a junction it heats or refrigerates the device due to the Peltier effect. In the non-local spin valve geometry, the interplay between both effects leads to a baseline resistance $R_{1}$ which is purely thermoelectric in nature.

The inclusion of Joule heating makes heat transport, and therefore also charge transport, nonlinear. This results in nonlinear voltage signals. While the linear response R$_{1}$ is often resulting from conventional Ohmic paths\cite{Johnson2}, the higher harmonic response are the result of thermoelectric effects. For example, the $R_{2}$ response can result from Joule heating of devices measured by a ferromagnet/non-magnetic thermocouple. If the typical bulk thermal dependencies $\gamma,\alpha$ of the Seebeck coefficient S(T)=$S_{0}(1+\gamma\Delta T)$ and conductivity $\sigma(T)=\sigma_{0}/(1+\alpha\Delta T)$ are used in the model, it is possible to explain even higher order nonlinearities observed in experiment\cite{Bakker}.

\subsection{Spin-Orbit Effects}

Whenever magnetic fields are applied or magnetism is involved in experiments, a large variety of spin-orbit effects may occur. These effects break the symmetry of the thermoelectric model. The most notable of these is the Hall effect which describes the generation of a voltage perpendicular to the plane made by an applied charge current and magnetic field. The observed transverse voltages generated in a ferromagnet are typically larger than expected from the ordinary Hall effect resulting from internal magnetic fields. Therefore, in ferromagnets, it is named the anomalous-Hall effect. The exact origin of this effect has been under continuous debate for over a century and seems to depend on the specific conditions of the experiment\cite{Nagaosa}.

The equation which describes the anomalous-Hall effect is obtained from Boltzmann transport theory\cite{Ashcroft} $\nabla V^{AH}=-\theta_{AH} \vec m\times J_{c}$ where $J_{AH}$ is the contribution to the current due to the anomalous-Hall effect. It is incorporated in the model using a skew symmetric conductivity:

\begin{equation}\label{eq2:AnHall}
\sigma(ij)=\sigma\left(\delta_{ij}-\theta_{AH}\sum_{k} \varepsilon_{ijk} m_{k} \right)
\end{equation}

\noindent Where $\theta_{AH}$ is the anomalous-Hall angle. Such contributions are often called spurious in spintronic devices since they can mimic spin valve signals\cite{Jedema}. However they can be modeled in spintronic devices to calculate their magnitude.

The Seebeck, Peltier and thermal conductivity coefficients have similar spin-orbit contributions. The effects which arise due to these contributions are named thermomagnetic effects. The effect arising from the Seebeck coefficient is named the anomalous-Nernst effect\footnote{Equivalently named the transverse Seebeck or Nernst-Ettingshausen I effect.} and describes the generation of a voltage perpendicular to the plane made by an applied heat current and magnetization. The contribution to the Peltier effect is named the anomalous-Ettinghausen effect\footnote{Equivalently named transverse Peltier or Nernst-Ettinghausen II effect.} while the contribution to the thermal conductivity is named the anomalous Righi-Leduc effect\footnote{Equivalently named the thermal-Hall effect}.

In addition, spin-orbit effects also determine that the conductivity of a ferromagnet is different by measuring it parallel or perpendicular to the magnetization of the ferromagnet. This is the anisotropic magnetoresistance effect and can be included in the model by adding an anisotropic contribution to the conductivity. The symmetric contribution then becomes\cite{Slachter2}:

\begin{equation}\label{eq2:AMR}
\sigma(ij)=\sigma_{\perp}\left(\delta_{ij}-R_{AMR} m_{i} m_{j} \right)
\end{equation}

\noindent Where $\delta_{ij}$ is the Kronecker delta and R$_{AMR}$ the coefficient governing anisotropic magnetoresistance, typically a few percent\cite{Costachespinpump}. By symmetry, we expect equivalent anisotropic contributions to the Seebeck, Peltier and thermal conductivity. However, specific measurements demonstrating these effects have not been reported to the authors knowledge.

We recently measured two of the previously discussed effects in a spin-caloritronic device\cite{Slachter2}. This device is best described as the thermal equivalent of a spin-valve structure (the magnetic heat valve) and will be discussed in more detail afterwards. Here, one of the ferromagnets was Joule heated to generate a heat current $Q$ through the spin-valve device. The temperature was measured on the second magnet using a thermocouple. Anisotropic magnetoresistive heating and anomalous-Nernst were found to dominate the measured voltage behavior. By virtue of a finite-element model, it was possible to determine magnetization angles and also the size of the anomalous-Nernst effect.

\section{Thermoelectricity and Spin}

Thermoelectricity extends charge transport theory and includes effects governed by the Seebeck, Peltier and thermal conductivity coefficients. The spin-transport model extends charge transport theory to include the spin-dependency of the conductivity and introduces the concept of spin-dependent voltages V$_{\uparrow,\downarrow}$. The model which extends charge transport theory to include both thermoelectricity and spin-transport is named the thermoelectric-spin model. It has been used for almost 50 years to describe thermoelectricity in ferromagnets\cite{Barnard} and more recently, to describe thermoelectricity of multilayered spin valves\cite{Gravier1,Gravier2,Hatami} and spin transport in ferromagnets\cite{Saitoh,Uchida2,Jaworski}.

The relevant physics and measurable voltages in devices can be calculated using finite-element modeling. The spin dependent voltages and temperature are the variables $\vec u$ = $\left(V_{\uparrow},V_{\downarrow},T\right)$ and the fluxes are determined by the spin-dependent charge currents and heat current $\vec J$ = $\left(J_{\uparrow},J_{\downarrow},Q\right)$. The conductance matrix now allows us to include spin-dependent Seebeck S$_{\uparrow,\downarrow}$ and Peltier coefficients $\Pi_{\uparrow,\downarrow}$ to describe not only the coupling between charge and heat transport, but also the coupling between spin and heat transport. The conductance matrix is given by:

\begin{equation}\label{eq2:condmatspinTE}
\bar c = \left(\begin{array}{ccc} \sigma_{\uparrow} & 0 & \sigma_{\uparrow}S_{\uparrow} \\ 0 & \sigma_{\downarrow} & \sigma_{\downarrow}S_{\downarrow} \\ \sigma_{\uparrow}\Pi_{\uparrow} & \sigma_{\downarrow}\Pi_{\downarrow} & k \end{array}\right)
\end{equation}

The conservation of charge currents remains unchanged. However, the Valet-Fert equation is altered because in the derivation thermoelectricity is disregarded. It is originally derived using particle conservation\cite{jedemathesis}:

\begin{equation}\label{eq2:particlecons}
\frac{1}{e}\nabla \cdot J_{\uparrow,\downarrow} = \mp \frac{n_{\uparrow}}{\tau_{\uparrow\downarrow}} \pm \frac{n_{\downarrow}}{\tau_{\downarrow\uparrow}}
\end{equation}

\noindent Here $\tau_{\uparrow\downarrow}$ represents the time for a spin up electron to flip its spin to spin down while $\tau_{\downarrow\uparrow}$ represents the time from a spin down electron to flip its spin to spin up. The excess electron densities are given by the Einstein relation $n_{\uparrow\downarrow}=N_{\uparrow\downarrow} e V_{\uparrow\downarrow}$, with $N_{\uparrow\downarrow}$ the spin dependent densities of states at the Fermi energy. In the thermoelectric-spin model, the spin-dependent charge currents $J_{\uparrow,\downarrow} = -\sigma_{\uparrow,\downarrow} \left(\nabla V + S_{\uparrow,\downarrow}\nabla T\right)$ additionally includes a temperature gradient as well as the spin-dependent Seebeck coefficients.

The Seebeck coefficient describes how the conductivity depends on energy and is described by the Mott formula. By virtue of the Einstein relation, the Seebeck coefficient is determined by the energy derivative of the density of states $\left(\frac{dN}{dE}\right)_{E_{F}}$ and the relaxation time $\left(\frac{d\tau}{dE}\right)_{E_{F}}$ at the Fermi energy. To develop an altered Valet-Fert equation, the energy dependence of the densities of states N$_{\uparrow,\downarrow}$ and relaxation times $\tau_{\uparrow\downarrow,\downarrow\uparrow}$ needs to be taken into account at the right side of Eq. \ref{eq2:particlecons}. While theoretically these contributions can be taken into account, in practice not much is known about these specific energy dependencies. For simplicity, we ignore such terms in the modeling and note that they can be responsible for small bulk source terms\cite{Saitoh,Slachter}. Conservation of charge, spin and heat currents are now taken directly from the individual thermoelectric and spin-transport models and produce the following source term:

\begin{equation}\label{eq2:TEspinsource}
\vec f = \left(\begin{array}{c} \frac{(1-P_{I}^{2})\sigma}{4\lambda^{2}}(V_{\uparrow}-V_{\downarrow})\\ -\frac{(1-P_{I}^{2})\sigma}{4\lambda^{2}}(V_{\uparrow}-V_{\downarrow}) \\ J_{\uparrow}^{2}/\sigma_{\uparrow} + J_{\downarrow}^{2}/\sigma_{\downarrow} \end{array}\right)
\end{equation}

\noindent Where we introduced the individual Joule heating of both spin channels $J_{\uparrow,\downarrow}^{2}/\sigma_{\uparrow,\downarrow}$.

In the spin-dependent charge transport model, the ferromagnetic/non-magnetic interface plays a crucial role in converting spin transport into charge transport and vice versa. Using this coupling between spin and charge, magnetic memory elements can be constructed. In the thermoelectric-spin model, the ferromagnetic/non-magnetic interface plays a similar role. In the following we will show that this interface can be used to convert heat transport into spin transport and also vice versa. The conversion of a heat current into a spin voltage depends solely on the spin-dependency of the Seebeck coefficient and is therefore named the spin-dependent Seebeck effect or by its more application oriented name: thermal spin injection. This was recently measured in a dedicated spin-caloritronic device\cite{Slachter}.

Here, we also propose a measurement scheme for the reciprocal effect. We show that when a spin current is injected into a ferromagnet a net heat flow develops, even in the absence of a charge current. This effect, named the spin-Peltier effect, depends solely on the spin-dependency of the Peltier coefficient.

\subsection{Spin-dependent Seebeck effect}

The spin-dependent current model dictates that when a charge current $J_{c}=J_{\uparrow}+J_{\downarrow}$ is sent through the bulk of a ferromagnet, a spin current $J_{s}=J_{\uparrow}-J_{\downarrow}$ accompanies it of which the size is determined by the conductivity polarization $J_{s}=P_{I} J_{c}$.

\begin{figure}[t]
\includegraphics[width=8.8cm,keepaspectratio=true]{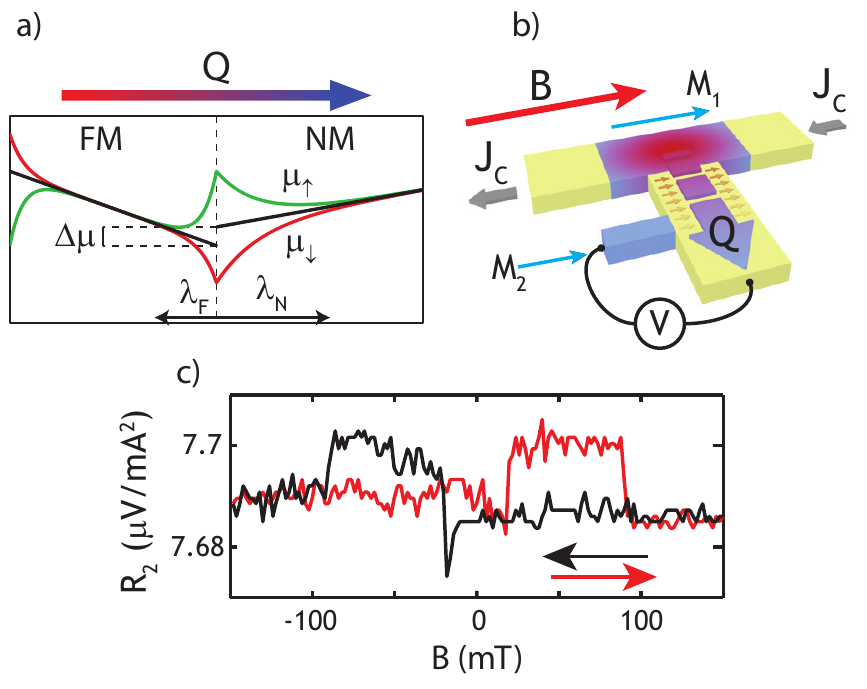}
\caption{\label{fig2:3} (Color online) Thermal spin injection, measurement scheme and measured result\cite{Slachter}. a) A heat current sent through a F/N interface generates a spin voltage $V_{s}\approx\lambda_{F}S_{s}\nabla T$ at the interface which extends a distance $\lambda_{N}$, $\lambda_{F}$ in the materials. b) Thermal spin injection can be measured by Joule heating FM$_{1}$ in a lateral spin valve. This generates a heat current $Q$ over the FM$_{1}$/NM interface which injects spins. The spin voltage is turned into a measurable voltage by the FM$_{2}$/NM interface. The size of thermal spin injection can be determined by selectively switching the magnetizations. c) Measurement result. A signal due to thermal spin injection is present on a large background caused by the measurement of Joule heating by the FM$_{2}$/NM thermocouple. In addition, small traces of anisotropic magnetoresistive heating effects as well as anomalous-Nernst effects are present. These can be seen by the small dip at the switching field of FM$_{1}$ and the offset in background voltage between large positive and negative magnetic fields (see also fig. 7).}
\end{figure}

In similar fashion, the thermoelectric-spin model dictates that when a heat current $Q$ is sent through the bulk of a ferromagnet in the absence of a charge current, a spin current $J_{s}=-\sigma_{F}(1-P^{2})S_{s}\nabla T/2$ flows, of which the size is determined by the spin-dependent Seebeck coefficient $S_{s} = S_{\uparrow}-S_{\downarrow} = P_{s}S$. Here $P_{s}$ is a fraction of the regular Seebeck coefficient $S$, defined in terms of the spin-dependent Seebeck coefficients as:

\begin{equation}\label{eq2:seebeckcoeff}
S=\frac{\sigma_{\uparrow}S_{\uparrow}+\sigma_{\downarrow}S_{\downarrow}}{\sigma_{\uparrow}+\sigma_{\downarrow}}
\end{equation}

In a non-magnetic material, both the conductivity polarization and the spin-dependent Seebeck coefficient are zero. When a charge current is sent through a ferromagnetic/non-magnetic interface, the discontinuity in bulk spin current creates a spin voltage at the interface and injects a net spin current in the non-magnetic material. The same situation occurs when a heat current in the absence of a charge current is sent through the interface.

This effect, the injection of spins in a non-magnetic material by a heat current, is named thermal spin injection or, equivalently, the spin-dependent Seebeck effect\cite{Slachter}. Since both electrical and thermal spin injection arise from the discontinuity of the bulk spin currents, both effects have similar behavior. For example, the spin voltage spreads an equal distance away from the interface and both effects suffer from the conductivity mismatch problem\cite{Schmidt} which strongly reduces spin injection in low conductivity materials such as semiconductors.

Thermal spin injection was recently demonstrated in a multiterminal lateral device\cite{Slachter}. In this device, a temperature gradient is applied to a F/N/F spin valve by Joule heating one of the ferromagnets with the help of a large charge current. The thermoelectrically generated spin voltage across the first ferromagnetic/non-magnetic interface is measured by a second ferromagnetic/non-magnetic interface which converts the spin voltage into a measurable voltage. A schematic picture of thermal spin injection and the used measurement scheme is shown in Fig. \ref{fig2:3}.

Because Joule heating scales quadratically with the applied charge current $I^{2}$, thermal spin injection results in a nonlinear spin dependent signal $R_{2}^{s} = R_{2}^{P}- R_{2}^{AP}$ where $R_{2}^{P}$ and $R_{2}^{AP}$ are the parallel and antiparallel contributions. The measured result is shown in Fig. \ref{fig2:3}c. The applied temperature gradient was very limited due to the relatively large lateral size and because electromigration prohibits heating in excess of 40K in this particular case.

In addition to thermal spin injection, we observe small traces of spin-orbit effects such as the anomalous-Nernst effect in the second ferromagnet and anisotropic magnetoresistive heating of the first ferromagnet. These effects have been more thoroughly examined in another device\cite{Slachter2}. These effects express themselves by a difference in background voltage for both parallel orientations and the observed small curve in voltage prior to the low field switch.

An application of spin currents lies in its ability to switch the (uniform) magnetization of a ferromagnet around its easy axis by means of spin-transfer torque\cite{Chappert}. This effect has a threshold in the spin current which needs to be injected in a small volume ferromagnet. Since electrical and thermal spin injection have the same physical origin, the discontinuity in bulk spin current, we may directly compare the critical temperature gradient in the ferromagnet needed to switch a F/N/F spin valve by spin transfer torque to the critical charge current density for this switching process. If the critical charge current density is known, we can calculate the critical temperature gradient which is needed for the switching process:

\begin{equation}\label{eq2:thermalspininjectiontreshhold}
\nabla T|_{crit} = \frac{2P_{I}}{\sigma P_{s}S (1-P_{I})} J_{crit}
\end{equation}

\noindent Here $J_{crit}$ is the threshold in charge current at which spin transfer torque switching takes place, $\sigma$ the ferromagnetic conductivity and $\nabla T|_{crit}$ the critical temperature gradient in the bulk ferromagnet. As an example, if we assume a critical charge current density J$_{crit} = 10^{11}$ A/m$^{2}$ for a permalloy ferromagnet, with the common (estimated) parameters P$_{I}=0.6$, P$_{s}=0.6$, $\sigma=4\cdot10^{6}$ S/m and $S=-20\mu V$ we find a critical temperature gradient of $\nabla T = 4\cdot10^{9}$ K/m. A typical F/N/F stack of 25 nm then switches at an applied temperature difference of 100 degrees. This process is known as thermal spin transfer torque, and recently evidence has been found for it\cite{Yu}.

\begin{figure}[!t]
\includegraphics[width=8.8cm,keepaspectratio=true]{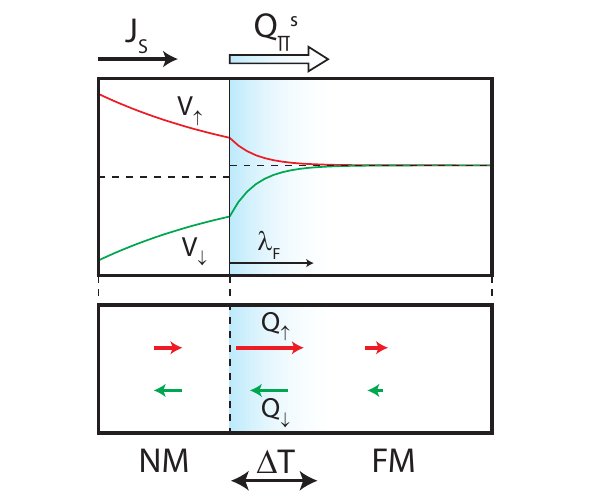}
\caption{\label{fig2:4} (Color online) Schematic representation of the Spin-Peltier effect. A spin current $J_{s}=J_{\uparrow}-J_{\downarrow}$ free of charge current ($J_{\uparrow}+J_{\downarrow}=0$) is injected from the non-magnetic side of the F/N junction into a ferromagnet $FM$. The top shows the resulting spin-dependent voltages, calculated using the 2-channel model. Despite the fact that no charge current is flowing through the junction, a net heat flow $Q_{\Pi}^{s}=\frac{1}{2}(\Pi_{\uparrow}-\Pi_{\downarrow}) J_{s}$ develops in the ferromagnet which quickly drops off due to the spin relaxation length $\lambda_{F}$. Depending on the sign of the spin current and the parallel/antiparallel alignment of the magnetization, net heat is transported from the non-magnetic material to the ferromagnet or vice versa. This creates a temperature difference $\Delta T$ between the bulk non-magnetic material and the bulk ferromagnet.}
\end{figure}

The ability to use finite element modeling should allow to engineer multiterminal F/N/F pillar devices which switch by thermal spin transfer torque. Such devices can combine the high polarization properties of pillar devices with the flexibility of lateral devices. By selectively heating the device, the effect can also be used to lower the effective threshold in electrical spin injection.

\subsection{Spin-Peltier effect}

The Onsager reciprocity relation dictates that when heat transport induces spin transport free of charge currents in a ferromagnet, the opposite can also occur. A pure spin current injected in a ferromagnet should induce net heat transport. This reciprocal effect is named the spin-Peltier effect.

\begin{figure}[!t]
\includegraphics[width=8.8cm,keepaspectratio=true]{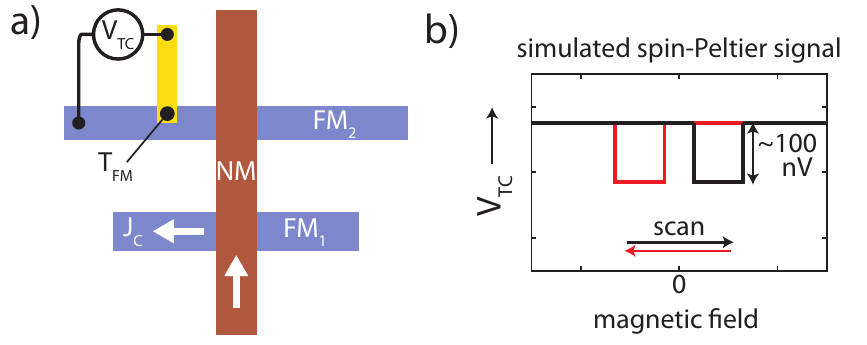}
\caption{\label{fig2:5} (Color online) Proposed non-local measurement scheme for the spin-Peltier effect. a) The non-local spin valve geometry allows to inject a pure spin current into ferromagnet $FM_{2}$ by sending a charge current $J_{c}$ across the $FM_{1}/NM$ interface. This spin current creates a temperature difference across the $FM_{2}/NM$ interface changing the local temperature $T_{F}$ of the ferromagnet. The effect is detected by converting the temperature to a voltage $V_{TC}=\Delta S_{FM} T_{F}$ using a thermocouple. Here $\Delta S_{FM}$ is the difference between Seebeck coefficients between the ferromagnet and the  non-magnetic material (yellow). The thermocouple measures the temperature $T_{F}$ of the ferromagnet and optionally, makes use of the large Seebeck coefficient of the ferromagnet itself. b) The simulated spin-Peltier signal. The resulting signal should have a background determined by Peltier cooling/heating at the $FM_{1}/NM$ interface and subsequent consequence to the temperature $T_{F}$. A small signal of $~100$ nV due to the spin-Peltier effect should arise which only depends on the parallel or antiparallel alignment of both magnetizations.}
\end{figure}

We illustrate this effect in Fig. \ref{fig2:4} by considering the F/N interface previously used. A pure spin current is injected from the non-magnetic side into the ferromagnet. When the spin current enters the ferromagnet it reduces in size at the spin relaxation length, for metals ranging from a few to tens of nanometers. The spin-dependent voltages which result from the spin current are sketched in the top part of Fig. \ref{fig2:4}. In the absence of charge currents, the heat current in the system due to the spin-Peltier effect $Q_{\Pi}$ can be deduced from the thermoelectric-spin model:

\begin{equation}\label{eq2:heattransportspinpeltier}
Q_{\Pi}=\frac{1}{2}(\Pi_{\uparrow}-\Pi_{\downarrow}) J_{s}
\end{equation}

In the non-magnetic material, $\Pi_{\uparrow}=\Pi_{\downarrow}$ and no net heat transport due to the spin-Peltier effect takes place. In the ferromagnet, the spin-Peltier coefficient, defined as $\Pi_{s}=\Pi_{\uparrow}-\Pi_{\downarrow}$ can be non-zero. Close to the interface, the spin current $\vec J_{s}$ is also non-zero. This induces heat transport due to the spin-Peltier effect which drops off in the ferromagnet at the spin relaxation length. As a result, a temperature difference $\Delta T$ develops between the bulk non-magnetic material and the ferromagnet.

The Thomson-Kelvin relation $\Pi=S T$ relates the conventional Seebeck and Peltier coefficients. This also holds for the individual spin species. From the recently found spin-dependent Seebeck coefficient $S_{s}$ we can calculate the spin-Peltier coefficient $\Pi_{s}=S_{s} T$ which can be used to estimate the effect.

We calculate the exact temperature difference by considering the total heat current $Q=\Pi_{s} J_{s}-k \nabla T$ in the ferromagnetic and non-magnetic regions. Like spin and charge currents, the heat current is continuous across the interface. If we ignore Joule heating due to spin currents, it is also continuous in the bulk of the non-magnetic and ferromagnetic parts and equal to $k_{i} \nabla T|_{i}$. Here the index $i$ denotes the region. The presence of the spin-Peltier effect induces an additional temperature gradient $\nabla T|_{\Pi} (x) = \frac{\Pi_{s}}{2 k_{F}}J_{s}(x)$ in the small ferromagnet region in which a sizeable spin current exists. The spin current drops off exponentially from the interface: $J_{s}(x)=J_{s}^{0} e^{-x/\lambda_{F}}$. Here $J_{s}^{0}$ is the spin current at the interface. If we integrate this additional temperature gradient over this region, we find the temperature difference between the bulk ferromagnet and non-magnetic material:

\begin{equation}\label{eq2:TempdiffSP}
\Delta T |_{F-N} = \frac{\Pi_{s}}{2 k_{F}} \lambda_{F} J_{s}^{0}
\end{equation}

This temperature difference depends solely on the spin-Peltier coefficient $\Pi_{s}$, the spin relaxation length $\lambda_{F}$ and the thermal conductivity $k_{F}$ of the ferromagnet. Its sign is determined by the sign of spin current and the spin-Peltier coefficient.

The non-local spin valve geometry is an ideal geometry to inject pure spin currents into a ferromagnet. The generated temperature difference over the interface can be detected by measuring the temperature of the ferromagnet in which the pure spin current is injected. This can be achieved by placing a thermocouple on the ferromagnet. This measurement geometry is illustrated in Fig. \ref{fig2:5}. The background voltage is then solely determined by the Peltier heating of the $FM_{1}/NM$ interface which injects the spin current and the subsequent measurement of the temperature by the thermocouple. The spin-Peltier signal then appears as a regular resistance $R_{1}$ which solely depends on the parallel or antiparallel alignment of both magnetizations.

For permalloy, all parameters are known\cite{Slachter} and we can estimate the temperature difference which can be created in this manner. At a realistic maximum pure spin current which can be injected into a permalloy ferromagnet in the non-local spin valve geometry (see the previous discussion below Fig. \ref{fig2:1}) we have $J_{s}^{0} = 10^{11}$ A/m$^{2}$. Using this value, we find a temperature difference of $\Delta T$ = 20 mK between the parallel and antiparallel orientation of the permalloy spin valve across the interface of the ferromagnet. A typical thermocouple which can be realized on a (lateral) ferromagnet\cite{Slachter2} has an efficiency of $\Delta S = 40 \mu V /K$. This results in a maximal spin-Peltier signal of 800 nV which is small but observable. Initial experiments show signs of the spin-Peltier effect, however, it is hard to distinguish it from small parasitic effects, for example, the pick-up of regular non-local spin valve signals by an uneven distribution of the spin-voltage at the detecting interface.

\section{Beyond Thermoelectricity and spin: The spin-dependent heat model}

In the thermoelectric-spin model a single electron temperature was introduced which holds for both spin species. The energy of the electrons is distributed in their respective bands according to a position-dependent Fermi-Dirac distribution $f_{\uparrow,\downarrow}(\epsilon,n_{\uparrow},T)$ with a spin-specific local density $n_{\uparrow,\downarrow}$ and local temperature $T$. This model assumes strong inelastic interaction within each spin species and also between them to obtain the required thermodynamic distribution. It is caused by electron-electron interaction or mediated by phonons through electron-phonon interaction.

At low temperatures inelastic scattering becomes weaker and this requirement does not hold. It was shown in the past that inelastic scattering can be weak on the scale of the spin relaxation length in non-magnetic metals\cite{Baselmans} at sub-4K temperatures where electron transport is still diffusive, limited by elastic scattering. Although in this situation it is hard to speak of electrons which are distributed according to a Fermi-Dirac distribution in their respective bands, it is still possible to describe thermal transport according to a diffusion equation. The temperature $T$ then represents the local average excess energy of electrons compared to the situation at zero Kelvin. In addition, the spin-dependent electron species also do not exchange energy with each other.

This requires the introduction of a spin-dependent heat model where both spin channels have their own heat current $Q_{\uparrow,\downarrow}$, thermal conductivity $k_{\uparrow,\downarrow}$ and spin-dependent temperature $T_{\uparrow,\downarrow}$. This opens up the possibility to demonstrate new thermal and, possibly, thermoelectric experiments in magnetoelectronic devices. Such a model has been first described by Heikkil\"a \textit{et al.}\cite{Heikkila}. We introduce a bulk diffusion model and calculate an example device.

The Wiedemann-Franz law describes the relation between charge and thermal conductivity. Both are dominated by electron transport in metals. Therefore, a spin polarization in the electrical conductance $P_{I}$ leads to a spin polarization in thermal conductance $P_{Q}$, similarly defined in terms of spin-dependent heat conductances as $P_{Q} = (k_{\uparrow}-k_{\downarrow}) / (k_{\uparrow}+k_{\downarrow})$. The model for spin-dependent electrical and thermal transport now includes the spin-dependent voltages and temperatures $\vec u = (V_{\uparrow},V_{\downarrow},T_{\uparrow},T_{\downarrow})$. The spin-dependent charge and heat currents $\vec J = (J_{\uparrow},J_{\downarrow},Q_{\uparrow},Q_{\downarrow})$ are determined through the 4x4 conductance matrix:

\begin{equation}\label{eq2:cmatrixspinheat}
\bar c = \left(\begin{array}{cccc} \sigma_{\uparrow} & 0 & \sigma_{\uparrow}S_{\uparrow} & 0 \\ 0 & \sigma_{\downarrow} & 0 & \sigma_{\downarrow}S_{\downarrow} \\ \sigma_{\uparrow}\Pi_{\uparrow} & 0 & k_{\uparrow} & 0 \\ 0 & \sigma_{\downarrow}\Pi_{\downarrow} & 0 & k_{\downarrow} \end{array}\right)
\end{equation}

Where the spin-dependent thermoelectric effects, represented by the coefficients $S_{\uparrow,\downarrow}$ and $\Pi_{\uparrow,\downarrow}$, are used in the relevant spin-dependent currents. The conservation of spin and charge currents remains the same, and therefore the components of the source term for the spin-dependent charge currents as well. It is straightforward to include the conservation of the total heat current $Q = Q_{\uparrow} + Q_{\downarrow}$ into its spin-dependent parts: the Joule heating of each channel $J_{\uparrow,\downarrow}^{2}/\sigma_{\uparrow,\downarrow}^{2}$ simply applies to the channels individually. We note here that strictly speaking, if there is no inelastic scattering, there is no Joule heating. However, any weak inelastic scattering does raise the average energy of the electron baths which allows Joule heating to be used in the model as a local source of heat.

Although it is beyond the scope of this article to derive the conservation of spin heat currents $Q_{s} = Q_{\uparrow} - Q_{\downarrow}$ from Boltzmann transport theory\cite{Tulaparkur2}, we may introduce a phenomenological relaxation analogue to the relaxation of the amount of spins themselves, represented by the Valet-Fert equation for spin voltage.

The difference in excess energy between both spin species is represented by the spin temperature $T_{s} = T_{\uparrow} - T_{\downarrow}$. In our model, we assume a thermal equivalent of the Valet-Fert equation $\nabla^{2}T_{s} = \frac{T_{s}}{\lambda_{Q}^{2}}$. Here $\lambda_{Q}$ is the relaxation length for the spin temperature. This relaxation length is not only limited by spin flip processes, but can also be limited due to inelastic scattering between both spin species, where energy is being exchanged between both spin species without flipping its spin. This results in the boundary condition for spin relaxation lengths $\lambda_{Q}\leq\lambda$. The spin relaxation lengths are equal whenever inelastic scattering is absent.

New source terms $\pm\frac{(1-P_{Q}^{2})k}{4\lambda_{Q}^{2}}(T_{\uparrow}-T_{\downarrow})$ are then added such that the conservation of spin heat is also included. This leads to the following source term in this model:

\begin{equation}\label{eq2:TEspinsource}
\vec f = \left(\begin{array}{c} \frac{(1-P_{I}^{2})\sigma}{4\lambda^{2}}(V_{\uparrow}-V_{\downarrow})\\ -\frac{(1-P_{I}^{2})\sigma}{4\lambda^{2}}(V_{\uparrow}-V_{\downarrow}) \\ \frac{(1-P_{Q}^{2})k}{4\lambda_{Q}^{2}}(T_{\uparrow}-T_{\downarrow}) + J_{\uparrow}^{2}/\sigma_{\uparrow} \\
-\frac{(1-P_{Q}^{2})k}{4\lambda_{Q}^{2}}(T_{\uparrow}-T_{\downarrow}) + J_{\downarrow}^{2}/\sigma_{\downarrow} \end{array}\right)
\end{equation}

The thermoelectric coefficients in this model typically scale with temperature and are very small at the temperatures where this model is applicable. For example, for many non-magnetic metals the Seebeck coefficient scales linearly with temperature such that typical Seebeck coefficients are in the order of 10 nV/K at Helium temperatures\cite{Barnard}. This also holds for typical ferromagnets such as Cobalt or permalloy. Therefore, we first explore the special properties of spin-dependent heat conduction itself and propose the thermal equivalent of the spin valve.

\subsection{Magnetic heat valve}

\begin{figure}[t]
\includegraphics[width=8.8cm,keepaspectratio=true]{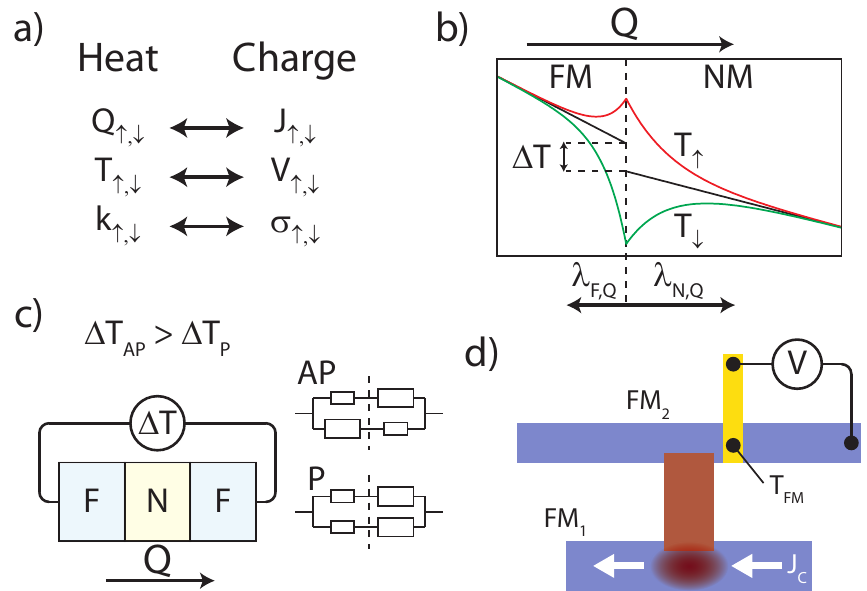}
\caption{\label{fig2:6} (Color online) The spin-dependent heat model. a) Equivalency between the coefficients of the spin-dependent heat and charge models when thermoelectricity and Joule heating is disregarded. b) A heat current $Q$ sent through the F-N interface creates a spin-temperature $T_{s}$ and spin-related temperature difference $\Delta T$. c) The F/N/F magnetic heat valve. A heat current sent through a F/N/F spin valve structure creates a temperature difference across it dependent on the parallel or antiparallel alignment of both magnetizations. d) A possible experimental realization of the F/N/F heat valve. By heat sinking one ferromagnet and Joule heating another, heat can be transported through the heat valve. The temperature of the second ferromagnet depends on the specific thermal resistance of the heat valve, determined through the parallel or antiparallel alignment of the magnetizations, and can be measured using a thermocouple.}
\end{figure}

If we disregard thermoelectricity as well as charge currents in magnetoelectronic devices, such that Joule heating is absent, spin-dependent charge and heat transport are each represented by an independent set of equations. The mathematical model introduced to describe spin-dependent heat transport is then identical to that which describes spin-dependent charge transport. The difference between the models is the size of the coefficients. The equivalence of the coefficients used in both models is depicted in Fig. \ref{fig2:6}a.

Consequently, concepts which are relevant in the spin-dependent charge transport model will have their equivalent in the spin-dependent heat transport model. A similar resistor model also applies\cite{Valet}. For example, consider a heat current $Q$ sent through a  F/N interface. This creates a difference in temperature between both spin species $T_{s}$ which relaxes in the materials with the spin heat relaxation length $\lambda_{Q}$. This is depicted in Fig. \ref{fig2:6}b. The size of the spin temperature at the interface can be directly deduced from the equivalently calculated spin voltage $V_{s}$ in the charge transport model\cite{vSon}:

\begin{equation}\label{eq2:heatvalvespintempFN}
\frac{T_{s}}{Q} = \frac{2P_{Q}R_{F,Q}R_{N,Q}}{R_{F,Q}+(1-P_{Q}^{2})R_{N,Q}}
\end{equation}

\noindent Where $R_{N,Q} = \frac{\lambda_{N,Q}}{k_{N}}$ and $R_{F,Q} = \frac{\lambda_{F,Q}}{k_{F}}$ are the equivalent thermal resistances determined by the spin heat relaxation lengths and the thermal conductivities of the materials. A spin related 'thermal resistance' $\Delta T = \frac{1}{2}P_{Q} T_{s}$ also develops across the interface.

There also exists a thermal equivalent of the electrical F/N/F spin valve. When a heat current $Q$ is sent through a F/N/F spin valve, a temperature difference $\Delta T$ develops across it, which depends on the parallel or antiparallel alignment of the magnetizations. We refer to this concept as the magnetic heat valve. It is depicted in Fig. \ref{fig2:6}c. In the electrical spin valve, a simple calculation\cite{Jedema} gives the difference between parallel and antiparallel resistance per unit area $R_{P}-R_{AP}= 2 P_{I}^{2} R_{F} R_{N}/ (R_{F}+R_{N}(1-P_{I}^{2}))$ whenever the distance $L$ between both ferromagnets is $L \ll \lambda_{N}$. Whenever $L \ll \lambda_{N,Q}$ we obtain the temperature difference between the parallel and antiparallel alignment in the magnetic heat valve:

\begin{equation}\label{eq2:heatvalveFN}
\frac{\Delta T_{P}-\Delta T_{AP}}{Q} = \frac{2 P_{Q}^{2} R_{F,Q} R_{N,Q}}{R_{F,Q} + R_{N,Q}(1-P_{Q}^{2})}
\end{equation}

\noindent As an example, let us consider a Py/Cu/Py heat valve in a 25 nm thick pillar stack where the non-magnetic metal is thin enough to satisfy the condition $L \ll \lambda_{N,Q}$. Whenever the spin valve is held at a total temperature where inelastic scattering is small but not negligible, say $\lambda_{Q}\approx \lambda/2$  and assume the estimated values $P_{Q} = 0.6, \lambda_{F} = 5$ nm, $\lambda_{N} = 1 \mu$m, $k_{Cu}$ = 300 W/m/K, $k_{Py}$ = 30 W/m/K an applied temperature gradient of 10K / 25 nm produces a significant temperature difference of $\approx$2K across the spin valve depending on the parallel or antiparallel alignment.

In a spin valve, it is possible to use Joule or laser heating to produce heat currents through a device\cite{Gravier1}. However, measuring a temperature difference across a device is non-trivial. Nevertheless, by measuring the absolute temperature at the second ferromagnet with the aid of a thermocouple, the process can be measured since the process does influence the net heat flow through the device. This experimental measurement technique is sketched in Fig. \ref{fig2:6}d. It is fairly non-trivial to use analytical solutions, as heat transport through a substrate and Joule heating itself are hard to calculate in a three dimensional geometry. However, by fitting the obtained measured voltages to those resulting from a finite-element model with varying geometry, it should be possible to extract useful coefficients, such as spin heat relaxation lengths at different temperatures.

\begin{figure}[t]
\includegraphics[width=8.8cm,keepaspectratio=true]{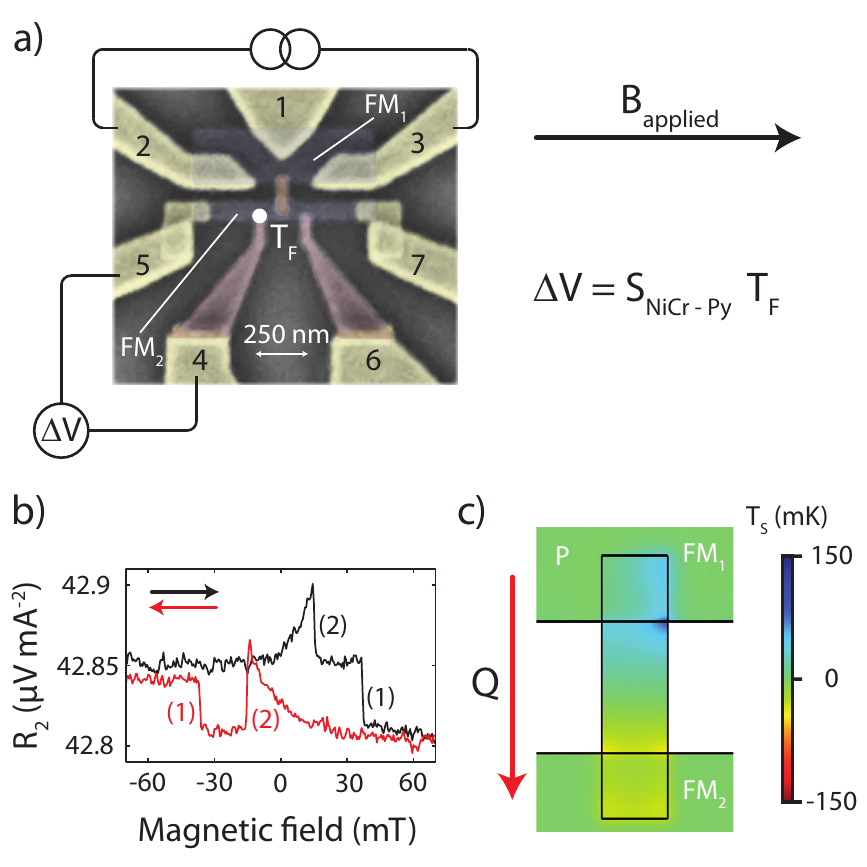}
\caption{\label{fig2:7} (Color online) Modeling of a fabricated device\cite{Slachter2} which potentially could show the magnetic heat valve effect. a) SEM figure and measurement geometry. Two permalloy ferromagnets (blue) are connected by a copper rectangle (brown). The temperature T$_{F}$ of the second ferromagnet is measured by a NiCr (contact 4,6) - Py thermocouple. b) Measurement at a typical current of 1 mA. No regular spin valve signal is present, which would be the result of the magnetic heat valve effect. The anomalous-Nernst (1) and anisotropic magnetoresistive heating (2) effects are present. c) Calculated spin temperature at I$_{1-2}$ = 1 mA zoomed at the copper rectangle calculated using the parameters $\lambda_{Py,Q}$ = 5nm, $\lambda_{Cu,Q}$ = 350 nm and P$_{Q}$ = 0.25.}
\end{figure}

This experimental measurement technique has been used previously at room temperature\cite{Slachter2}. Here, a Py/Cu/Py spin valve was used and a Py/NiCr thermocouple. A SEM picture of the device is shown with a typical measurement in Fig. \ref{fig2:7}. The system was modeled with a regular thermoelectric model which showed that a maximum heat current of $Q\approx10^{9}$ W/m$^{2}$ can be achieved at which the temperature difference across the spin valve is $\approx$2K with a used charge current of $I_{c}$ = 2 mA.

Although at room temperature, regular spin-orbit effects such as the anomalous-Nernst (1) and anisotropic magnetoresistance (2) are dominant, we can use this sample to demonstrate a calculation of the spin-dependent heat model. Using the additional parameters shown in Fig. \ref{fig2:7}, and assuming no inelastic scattering ($\lambda_{Q} = \lambda$) and $P_{Q}=P_{I}$, we calculate a temperature difference $\Delta T_{P}-\Delta T_{AP}$ = 8 mK due to the heat valve effect which leads to a response R$_{2}^{s}$ = 19.6 nV/mA$^{2}$ when the thermocouple is measured. The temperature difference is 10 times lower than the value calculated from Eq. \ref{eq2:heatvalveFN} with $\nabla T|_{F}\approx3\cdot10^{7}$ K/m and is due to spin relaxation in the non-magnetic material. With a noise level of $\approx$5 nV/mA$^{2}$ the heat valve effect should be observable if inelastic scattering between both spin species is absent. The absence of a spin heat signal above the noise level shows that inelastic scattering is strong enough such that we find the requirement $\lambda_{Q} < \frac{1}{2}\lambda$ valid at room temperature. When $\lambda_{Q} = \frac{1}{2}\lambda$ the calculated signal is approximately identical to the noise level.

\subsection{Thermoelectricity and spin-dependent heat}

The spin-dependent heat model becomes relevant when inelastic interaction between the spin species is weak, which occurs at low temperatures. Thermoelectric effects are small at these temperatures, which is why thus far we did not consider the connection between the spin-dependent thermoelectric effects and the effects due to spin-dependent heat.

However, prospects in fabrication which connects the flexibility of a multiterminal lateral device design with the high signals observed in pillar structures and the low noise experiments associated with a low operating temperature should increase the observability of the effects so far considered. We may then also consider the higher order effects related to this connection.

Whenever a charge current $J_{c}$ is sent through a ferromagnet a spin heat current $Q_{s} = \Pi_{s} J_{c}$ also flows, determined by the spin-Peltier coefficient $\Pi_{s}$. Here we assumed $V_{\uparrow}=V_{\downarrow}$. Similar to the case of electrical spin injection and thermal spin injection, when this charge current is sent through a F/N interface, this creates a spin temperature $T_{s}^{0}$ at the interface which relaxes in the respective materials at the respective spin heat relaxation lengths $\lambda_{Q}$.  We propose to name this effect the thermal spin-Peltier effect. The size of the effect is given by Eq. \ref{eq2:heatvalvespintempFN} with the source of the spin heat current in the bulk $Q_{s} = P_{Q}Q$ substituted by that due to this effect $Q_{s} = \Pi_{s} J_{c}$. We note that this ignores the generation of a spin voltage at the interface which by ordinary thermoelectric effects is converted to a spin temperature.

Although at low temperatures $\Pi_{s}$ can be very small, the maximum charge current, for small devices typically limited by electromigration ($J_{c}^{max}$ = 10$^{12}$ A/m$^{2}$), is often larger than the typical heat current, which might render this effect more efficient to generate a spin temperature in a non-magnetic material than the previously described effect in the magnetic heat valve\footnote{In practice, it is necessary to separate the creation and detection of a spin temperature in a lateral device in order to prove the effect is due to the spin temperature only. Otherwise, such effects can be mistaken for a spin-current injection related voltage spread over the detector interface. This geometry requires additional contacts which increases the distance between injection and detection and reduces the desired observable effect.}.

The Onsager reciprocal effect can also occur in this model. Whenever a spin heat current $Q_{s}$ is injected into a ferromagnet at a F/N interface this creates a voltage difference $\Delta V |_{F-N}$ between the bulk non-magnetic material and ferromagnet. We propose to name this the thermal spin-dependent Seebeck effect. The calculation of this voltage goes similar to the calculation of the temperature difference $\Delta T$ in the spin-Peltier effect. Checking the symmetry between the spin-dependent charge and heat models we can directly substitute the various coefficients in Eq. \ref{eq2:TempdiffSP} to obtain the induced voltage difference over the interface:

\begin{equation}\label{eq2:VoltdiffSP2}
\Delta V |_{F-N} = \frac{S_{s}}{2 k_{F}} \lambda_{F,Q} Q_{s}^{0}
\end{equation}

It is possible to measure this voltage difference directly over an interface in a multiterminal non-local device. However, it requires a source of pure spin heat current $Q_{s}$ in which preferably a charge-related spin current $J_{s}$ is absent. This is a situation difficult to achieve. However, if both sources of $Q_{s}$ and $J_{s}$ scale differently with applied charge current it is perhaps possible to distinguish between the generation of a voltage over a F/N interface due to charge-releated spin currents and those due to spin heat currents in a suitably designed experiment.

\section{Discussion}

Throughout this article we have considered transparent (Ohmic) interfaces and collinear magnetic systems. In this case, the spin-dependent charge and heat currents are continuous across the interfaces and are scalar quantities. Past experiments show that whenever oxide layers are formed at interfaces the spin-dependent effects can be greatly enhanced\cite{Parkin}. Furthermore, a non-collinear system is required to describe important applications of spin-dependent transport such as the spin-torque-oscillator\cite{Tsoi,Kiselev,Krivorotov,Kaka,Houssameddine} or spin-transfer-torque magnetic memory\cite{Chappert,Albert,Yang}. In these cases the diffusion theory developed here is not sufficient to describe the relevant processes. Instead, it should be described by a more general theory which includes spin-dependent tunneling and a 3-dimensional spin vector, for example the magnetoelectronic circuit theory\cite{BrataasNCME,Tserkovnyakrevmodphys}. Heikkil\"a, Hatami and coworkers have previously developed such a theory\cite{Heikkila,Hatami} in explaining thermal spin-transfer torque and spin-dependent heat.

In such a theory, the transport of electrons across interfaces is described by a 4x4 conductance matrix $\bar G$, which relates the total flux $\bar J=(J_{c},\vec J_{s},Q,\vec Q_{s})$ at the interface to the variables at both sides $\bar u^{i}=(V_{c}^{i},\vec V_{s}^{i},T^{i},\vec T_{s}^{i})$ (i=F,N) by $\bar J = \bar G (\bar u^{F} - \bar u^{N})$.

Initial calculations on the conductance matrix for F/N interfaces have been carried out by Hatami and coworkers\cite{Hatami} who in this framework calculated thermal spin-transfer-torque for various ferromagnet/non-magnetic interfaces.

In the extended magnetoelectronic circuit theory they have introduced, all spin-dependent physics such as electrical or thermal spin injection is fully determined by the elements of the conductance matrix instead of the previously defined bulk spin-polarized parameters $P_{I}, P_{S}, P_{\Pi}, P_{Q}$ of the ferromagnet. The physical origin in both theories is principally different; diffusion theory relates spin-dependent effects occurring at the interface to bulk ferromagnetic parameters (which in itself is only relevant very close to the interface) and the magnetoelectronic circuit theory describes the physics to occur due to quantum mechanical tunneling of spin states. However, the effects they describe are the same. For example, thermal spin injection still describes the injection of spins in a non-magnetic metal due a heat flow over a ferromagnetic/non-magnetic interface. Dependent on the application, either the more simple diffusive theory developed here can be used or one needs to refer to the full circuit theory.

\section{Summary}
We have developed a diffusive theory for spin-dependent charge and heat conduction which includes spin-orbit effects. Finite-element methods were used to model several experiments from literature where several parameters of this model were quantified. Electrical spin injection, the spin-Hall angle of platinum and thermal spin injection were calculated. Also, new experiments were proposed which should demonstrate the spin-Peltier effect and a lower limit was given in an experiment which failed to demonstrate the magnetic heat valve.
\section{Acknowledgement}
We would like to acknowledge J.G. Holstein, B. Wolfs and S. Bakker for technical assistance and F.K. Dejene for critically reading the manuscript. This work was financed by the European EC Contract IST-033749 'DynaMax' and the 'Stichting voor Fundamenteel Onderzoek der Materie' (FOM).


\end{document}